\documentclass[runningheads]{llncs}
\usepackage{graphicx}
\usepackage[colorinlistoftodos]{todonotes}
\usepackage{tabularx}

\newcolumntype{C}[1]{>{\centering\arraybackslash}p{#1}} 

\begin{document}
\title{Strategically using Applied Machine Learning for Accessibility Documentation in the Built Environment}
%
%
\author{Marvin Lange\inst{1} \and
Reuben Kirkham\inst{2} \and
Benjamin Tannert\inst{3}}
\authorrunning{M. Lange et al.}
%
\institute{University of Bremen, Bibliothekstr. 1, 28359 Bremen, Germany
\email{marvin4@uni-bremen.de}\\
\and  Monash University, Wellington Rd, Clayton VIC 3800, Australia
\email{reuben.kirkham@monash.edu}
\and City University of Applied Sciences Bremen, Flughafenallee 10, 28199 Bremen, Germany
\email{benjamin.tannert@hs-bremen.de}}
\maketitle              
\begin{abstract}
There has been a considerable amount of research aimed at automating the documentation of accessibility in the built environment. Yet so far, there has been no fully automatic system that has been shown to reliably document surface quality barriers in the built environment in real-time. This is a mixed problem of HCI and applied machine learning, requiring the careful use of applied machine learning to address the real-world concern of practical documentation. To address this challenge, we offer a framework for designing applied machine learning approaches aimed at documenting the (in)accessibility of the built environment. This framework is designed to take into account the real-world picture, recognizing that the design of any accessibility documentation system has to take into account a range of factors that are not usually considered in machine learning research. We then apply this framework in a case study, illustrating an approach which can obtain a f-ratio of 0.952 in the best-case scenario.

\end{abstract}
\section{Introduction}
According to the World Health Organization more than one billion people (15\% of the global population) have a disability \cite{2011World}. For many of these people, mobility remains challenging due to accessibility barriers in the built environment \cite{Clarke2008Mobility,Harpur2011Time,Meyers2002Barriers}. For example, stepped surfaces are inaccessible to any wheelchair user, whilst inappropriate surfaces like cobblestones can be a barrier for most people with mobility impairments. Because of these accessibility barriers, many people with disabilities avoid or limit their travel outdoors, which often excludes them from wider society, employment opportunities and limiting their quality of life \cite{Hara2016design,Rodger2016Technology}.

The mapping of inaccessibility in built environments is a form of empowerment in this context \cite{Rodger2016Technology}. The provision of comprehensive documentation of (in)accessibility is potentially the cornerstone for a growing accessible built environment and the disposal of existing barriers \cite{Froehlich2019Grand}. There is an emerging program of research aimed at using technology to assist with making this documentation process more effective, be it via automatic approaches relying on a range of sensors such as acceleration sensors or depth cameras \cite{Holloway2013micro-level,Iwasawa2016Combining,Mourcou2013Wegoto:}, direct inspections, done via crowdsourcing \cite{Hara2013Combining,Saha2019Project,Venues} or hybrid approaches that combine both methods \cite{Hara2013Combining,Hara2014Tohme:}.

At present, the necessary information about the accessibility of sidewalks is on one hand rarely available, and the limited information that is available is often unreliable or inaccurate \cite{Froehlich2019Grand}. Routing systems for wheelchair users regularly present routes containing unnecessary detours and/or inaccessible routes, when the shortest accessible route is needed \cite{Tannert2019Analyzing,Tannert2018Disabled}. Indeed, a cost benefit analysis even showed that existing systems in use are worse than using plain Google Maps \cite{Tannert2019Analyzing}. One aspect that leads to these errors are missing information or misinformation about the type of surface \cite{Tannert2019Analyzing,Tannert2018Disabled}. This basic knowledge about the surface is very important both for people with disabilities but also for elderly people who have age related impairments \cite{Kerr2012Role}.

We make two important contributions in this paper. First, we outline a decision-making framework for designing accessibility \textbf{documentation} systems (that could ultimately be used in a navigation or other system) based on a careful use of machine learning algorithms: the framework addresses the particular challenges that arise when addressing this unique context. This includes human rights and accessibility related considerations. Second, we then provide a proof-of-concept implementation of that framework to a particular scenario, applying it to the concern of surface classification in Europe. Our implementation has three distinguishing features (i) it involves a carefully chosen set of classes that can be practically recognized, (ii) it does not rely on inertial sensors or the motion of the person taking the records and (iii) focusses upon street segments, rather than trying to classify individual frames. Our evaluation shows that this approach is highly effective, obtaining an f-ratio of 0.952 in a best-case scenario.

\section{Background}
\subsection{Benefits of Documenting Accessibility in the Built Environment}
To understand the accessibility documentation problem, it is necessary to start with the purpose of documenting inaccessibility in the built environment. The overarching issue is one of human experience: the built environment has a multiplicity of physical barriers for people with disabilities. In some cases, this can have a serious negative impact on their day-to-day life, reducing mobility, access to the workplace, social opportunities and often increasing the level of care worker support that otherwise would not be needed if a location was accessible \cite{Clarke2008Mobility,Gharebaghi2018Role,Norgate2012Accessibility}. Effective documentation offers the following potential improvements on this situation:

\subsubsection{Navigation and Route-finding.} Existing navigation tools do not presently provide a reliable or effective means for people with mobility impairments to traverse the built environment \cite{Tannert2019Analyzing}, with the result of considerable (and often distressing) inconvenience for many people with mobility impairments \cite{Hara2015Improving}. The problem is the absence of underlying data, with existing maps failing to provide reliable documentation of accessibility barriers \cite{Froehlich2019Grand}, however the inaccessibility of a given system may also mean that some journeys are impossible \cite{kirkham_using_2021}. A route-finding system would need to identify for a given user, in line with their unique requirements, what route to undertake, with sufficient accuracy as to offer better routes than existing routing tools. Notably, offering better routes than what has gone before might not be a high-bar in practice, not least because Google Maps for pedestrians is currently the best performing system, despite having no specialist features for computing accessible routes \cite{Tannert2019Analyzing}: the simulations in \cite{kirkham_using_2021} now provide strong evidence for the feasibility of such an approach. These limitations are perhaps unsurprising, given the limited amount of work that focusses on navigation for people with mobility impairments (compared to vision impairments) \cite{Gupta2020Towards}.

\subsubsection{Determining Which Regions to Travel to.} People with disabilities have the opportunity to make choices, such as where to go on holiday, or what parts of a city to explore. Accurate knowledge of the general level of (in)accessibility in a given location, and thus the relative risk of not being able to effectively access it are an important part of making an informed choice: either on where to travel, or whether support is needed (e.g., a relative, or a care worker). In effect, this would provide a broad-brush assessment of a location’s level of accessibility, thus enabling a more informed choice, without having to rely upon a specialist travel agency (with the considerable expense and limited choice of locations this entails \cite{Bowtell2015Assessing}), or paying for extra support (from a limited budget) ‘just in case’ there is a problem.

\subsubsection{Determining and Influencing Change.} Presently, planners lack effective details of the built environment so they can optimize which physical barriers to address \cite{Kirkham2017WheelieMap:}. Given limited public resources, better accessibility mapping would allow for the built environment to improve more rapidly, by more optimal targeting (and prioritization) of physical improvements, including maintenance work \cite{Gharebaghi2018Role}. A general picture that allowed the comparison between municipalities and nation states would also enable more effective supervision of performance (by human rights NGO’s, the UN and disability advocates alike) and thereby furthering Article 31 of the UN CRPD (which requires states to collect data to formulate and implement policies, as well as to demonstrate compliance with that convention) \cite{Convention}.

\subsubsection{Measuring the Effect of the Built Environment.} Fitness tracking has been recognized as an important concern, especially for people with mobility impairments \cite{Carrington2015"But}. Whilst it is now possible to effectively track the number of strokes with a fitness tracker \cite{FastCompany2016How}, tracking \emph{where} someone has been can provide more detailed performance information: this is because it is more (or less) effort to traverse different types of surfaces \cite{Meyers2002Barriers}. For someone learning how to use a manual wheelchair, this could also be an indication of rehabilitation progress and/or increased skill/confidence in using the chair: notably measuring real-world skill and performance is an important part of wheelchair skills training \cite{Kirby2002Wheelchair,Kirby2016Wheelchair} (as well as separately, the assessment of disability benefits which focus on the level of functional impairment \cite{Watson2020PIP}). Separately, rough surfaces increase harmful whole-body vibration, with a long-term negative effect upon health \cite{Garcia-Mendez2013Health,Wolf2005Vibration}, so tracking the quality could also be useful for longitudinal healthcare studies of wheelchair users.

\subsection{Existing Approaches towards Documenting (In)Accessibility in the Built Environment}

\subsubsection{An Ideal Accessibility Measurement System.} One way to understand the limitations of a measurement system is to compare it against an ideal system. An ideal system would \emph{accurately} capture \emph{all} relevant categories of accessibility information, be it the \emph{presence or absence of the appropriate affordances} (such as dropped curbs \cite{Kirkham2017WheelieMap:}, audible street crossings \cite{Ahmetovic2015Zebra,Kirkham2015Can}) or the \emph{quality of the surface} itself (e.g., does it contain trip hazards \cite{Chen2011Wheelchair-related}, or cause dam-aging whole-body vibration \cite{Bowtell2015Assessing,Rice2018Quality}). It should also capture \emph{temporary or transient} barriers (e.g., a street that is partially obstructed for blind people because bins are placed on the street \cite{Atkin2010Sight,Norgate2012Accessibility}), as well as an \emph{up-to-date} picture of the environment itself.  Finally, it should be \emph{economically efficient} enough to ensure \emph{complete coverage}, so that it can be used effectively in a navigation system \cite{Froehlich2019Grand,Tannert2019Analyzing}.

\subsubsection{Manual Expert-Driven Documentation.} The most notable expert driven system is AccessAble (formerly known as Disabled Go) \cite{AccessAble}. This involves a rigorous and expensive process of direct documentation by an expert of every feature in the built environment within a confined area (e.g., the local high street), performed by an expert assessor on behalf of the local authority. PhotoRoute \cite{PhotoRoute} serves a similar, but discrete function: a venue can request a photo illustrated accessible route from an important landmark (e.g., a Train Station) to enable attendees to follow the most appropriate route. Whilst expert-driven systems are effective for the limited routes and locations that they actually cover, they have the limitation of sparse coverage caused by the underlying expense of adopting them (these exercises are notoriously “laborious and time consuming” \cite{Froehlich2019Grand}). The overall effect they simply do not scale to provide adequate coverage, with the exception of confined (but important) situations.

\subsubsection{Semi-Automated and Crowdsourcing Approaches.} There has been a considerable focus on crowdsourced approaches with the aim of obtaining scalability. Perhaps the most successful has been Project Sidewalk \cite{Saha2019Project}, which has managed to cover an entire urban area using Google Street View data, however, there were challenges with annotator accuracy in respect of certain data categories. Other semi-automated approaches have focused on accelerating the video-documentation of barriers in the town-planning setting, such as Wheelie-Map \cite{Kirkham2017WheelieMap:}. For providing landmarks for people with visual impairments, this has been successfully crowdsourced in the context of public transport \cite{Hara2015Improving}.

Less-structured ‘geocrowdsourcing’ approaches have encountered significant problems in respect of their ability to accurately document barriers [54], whilst lacking effective user engagement and thus coverage \cite{Froehlich2019Grand}.  A notable ‘real world’ exception is the widely used FixMyStreet \cite{FixMyStreet}, which is a \emph{MySociety} project aimed at general problems in the local community, but also serves as a platform for re-porting some accessibility related problems (e.g., potholes). At present, there is no crowdsourced or semi-automated system that works in real-time for obtaining standardized reports of inaccessibility, nor one that does not rely on Google Street View (which can be many years out of date and is effectively banned in many countries, e.g., Germany \cite{Meinke2018Kamera-Autos}).

\subsubsection{Automated Approaches.} In an ideal world, the most effective approach towards accessibility documentation would be an automated one, given the relatively lower cost. There are numerous works that claim to offer a viable system for using inertial sensors mounted on mobility aids (e.g., wheelchairs) to measure the accessibility of the built environment \cite{Iwasawa2015Toward,Iwasawa2012Life-logging,Iwasawa2016Combining,Kurauchi2019Barrier,Yairi2019Estimating}. Unfortunately the evaluations that are said support these claims are based on 'toy problems' in respect of the data being collected in a laboratory rather than in the real world (meaning the findings are unlike to apply there  \cite{Poppe2007Evaluating}. Furthermore, they fail to use the appropriate leave-one-out metric of evaluation \cite{Hammerla2015Let's} (in some cases, participants were left out, but they all followed the same route, which is also a serious error). Another problem is that there are no performance metrics used that relate directly to problems in accessible navigation: typically F-scores or precision and recall of instances were used, which does not directly relate to problems of concern: whilst there are alternative metrics (e.g., as in \cite{Ward2011Performance}), these do not apply to accessible navigation.

In respect of work outside of inertial sensing, there has been some modest success. Some work has explored using aggregate GPS mapping to indicate where people have \textbf{not} gone \cite{Mora2017comprehensive,Palazzi2010Path}, but that overlooks surface quality, and overlooks the subjective nature of accessibility barriers. Computer vision has been success-fully used to detect tactile paving \cite{Ghilardi2016new}, in hybrid approaches aimed at supporting annotation \cite{Hara2013Combining,Hara2014Tohme:} and in identifying street crossings \cite{Ahmetovic2015Zebra,Ahmetovic2017Mind}. Deep learning approaches have been recently used to identify street features and accessibility barriers in the United States, yet there remains substantial challenges in respect of achieving sufficient accuracy for a navigation system, with an F-score of around 0.8 being the present state of the art \cite{Weld2019Deep}.

\section{A Practical Framework for Designing a Machine Learning based Accessibility Documentation System}
Automating accessibility documentation is in effect an applied machine learning problem. However, this problem has unusual and important qualities that present particular challenges, meaning it cannot be treated as a usual human activity recognition or documentation problem. For instance, there are substantial difficulties in determining what to classify (given the breadth of disabilities involved \cite{Gupta2020Towards}), there are substantial ‘grand challenges’ (such as the need for economic efficiency) \cite{Froehlich2019Grand} and there is also the need to design documentation tools that are themselves accessible for disabled people. We distill the design challenge into five considerations that collectively constitute a \textbf{framework} for designing a machine learning approach in this setting.  Whilst many of these observations are arguably routine (or even obvious), they all reflect concerns that are most often overlooked or not properly addressed (and which have important disability specific qualities), hence the need for this exercise.
\subsection{Issue 1: What to classify and how to classify it?}
Machine learning operates on the basis of data: a set of labelled data is used to train (and then test) an algorithm ‘learned’ through this process. In practice, this involves consistently annotating a dataset so that each item is labeled with one or more classes or categories. There is limited agreement amongst disabled people as to what constitutes an accessibility barrier  or which barriers are most important \cite{Kirkham2017WheelieMap:}, making such categorization controversial. This is especially so given that existing research on accessible navigation has been skewed away from certain groups \cite{Gupta2020Towards}, thus there is a need to ensure that a wider range of measures of potential inaccessibility are provided going forwards. The choice of categories also needs to work effectively from an engineering perspective, which is challenging enough at the best of times \cite{Nguyen-Dinh2013Tagging,Reyes-Ortiz2015Transition}. This means there is a need to carefully select and then justify an appropriate set of classes in respect of the challenging disability-specific scenario, whilst ensuring they are well defined enough to enable a robust division between testing and training data.  

\subsection{Issue 2: How to measure performance (and success)?}
Measuring performance involves dividing data into testing and training datasets, before training an algorithm on the training data, and then testing the resulting algorithm on the testing data. For this to be a fair comparison, there needs to be a proper division between testing and training data. In the scenario of activity classification, the widely used ten-fold random cross-validation is inappropriate due to the way it mixes these two categories together and inflates performance: instead, \emph{a leave one out strategy} should be used to avoid this problem \cite{Hammerla2015Let's} With respect to inertial sensing for accessibility documentation, not using a leave one out approach has been shown to vastly inflate reported performance \cite{Mascetti2020SmartWheels:}: nevertheless, this inappropriate practice remains widely used.

In machine learning research, different algorithms are benchmarked using standardized datasets in order to compare their performance \cite{Stallkamp2012Man}. Yet, at present, there is no such dataset for accessibility recognition in the built environment (nor given the diversity of approaches, is one desirable or possible), representing an-other substantial problem, in that it is not possible to easily compare systems and works in this domain. Another issue in measuring performance is to use domain-appropriate metrics \cite{Ward2011Performance}, that indicate the real world implications of the system if it were to be translated into practice: yet there are no domain-appropriate metrics for accessibility mapping. Activity recognition systems can be particularly prone to bias in respect of certain types of disabled people, in part due to limitations with testing and training data (and how this is evaluated), which is an issue that has only just begun to be addressed \cite{Guo2019Toward,Kurauchi2019Barrier,Trewin2018AI}. Thus measuring performance is one of the biggest concerns in designing an accessibility documentation system, especially as this determines what will be deployed in practice (especially when experiments are conducted in order to choose which approach to use, as is usual in machine learning research \cite{Drummond2010Warning:}).

\subsection{Issue 3: What measurement apparatus to use?}
The choice of sensing apparatus underpins the design of a system: it is the source of information for classification. This involves deciding between what sensors to use (e.g., accelerometer, gyroscope, camera) and how/where to mount them, be it to a person, wheelchair or any other device that is capable of movement or being propelled. One important disability related challenge is the diversity of motion (arising from each individual mobility impairment): consideration has to also be given to the population who propel a device or to whom a sensor is attached. For example, different wheelchair users propel their chairs in different ways, and each wheelchair is highly customised to the specific anatomical requirements of its user \cite{Trefler1991Prescription}, thus a source of variability that risks any algorithm not generalizing across wheelchair users, if it is dependent on how they move (approaches such as static cameras are less problematic in this regard). Thus, the wider concern of ‘bias’ and ‘fairness’ in AI, is also a prominent concern in respect of the choice of measurement apparatus \cite{Guo2019Toward,Morris2020AI,Trewin2018AI}. Moreover, there is need to ensure that any means of affixing and operating the hardware is physically accessible in and of itself (typically commodity camera clamps have been used, which works well for manual wheelchair users, but might not be so effective in other circumstances \cite{Kirkham2017WheelieMap:}). The choice of measurement apparatus is therefore particularly complicated in respect of people with mobility impairments. As such, the choice of measurement apparatus must work practically for the people who have to use it (taking into account any relevant disabilities), as well as not depending on the characteristics of the individual user.

\subsection{Issue 4: What (machine learning) algorithm to use?}
Machine learning involves representing data by a list of features, which is a set of numbers that describes each element to be classified. When deployed, these features are then fed into a classifier to produce a prediction, or label. There are multifarious ways of providing features, including statistical means and moments, as well as specialized functions and rules \cite{Hammerla2013On}: in some cases, as with neural net-works, the pixels of an image can be submitted directly as features to an input layer \cite{Ordonez2016Deep}. The machine learning algorithm will learn the appropriate decision rule based on the data provided via its feature representation. Thus, there are two choices to make: the first is the feature representation, whilst the second is the machine learning classifier itself (e.g., KNN, SVM), although in some cases (e.g., Deep Learning) these issues somewhat shade into each other.  This combination can have practical issues, for example different features and classifiers are more robust to different types of error, whilst there can be considerable effects on battery life and speed of prediction depending on whether features are handcrafted and if certain forms of classifiers are used \cite{Haresamudram2019On}. 

Typically, the choice of classifier is decided experimentally. The difficulty in respect of disability arises this way: given the variety of settings and different impairments types, there is a need to particularly emphasize the robustness of the algorithm to signal noise and the diversity of different approaches. This can also require using a care-fully constructed training/testing set that properly takes this into account. At the same time, there can be wider considerations that go beyond performance: for instance, there is a need in some cases to provide transparency of decision making, which itself varies across classifiers. 

\subsection{Issue 5: How to address legal and economic considerations?}
A practical system for measuring the accessibility of the built environment has to work in the real world: the nature of disability poses challenges, this time from a societal perspective. For disability, the primary concern is an economic one, with accessibility mapping systems having wholly insufficient coverage due to the expense of measuring and documenting accessibility barriers \cite{Froehlich2019Grand}. So a successful system must have a clear economic model and roadmap that has sufficient cover-age: from a practical point of view, it should be suitably efficient and inexpensive to obtain the required data. Another consideration is legal: the main barrier is privacy concerns (a barrier that has tripped up systems such as Google Street View in many countries, especially in Europe, where there are human rights considerations). However, the law also presents opportunities, as it places obligations on organizations, potentially including providing supporting infrastructure \cite{Kirkham2015Can} under disability human rights law. Designers should consider how to use legal obligations (especially the duty to make reasonable accommodations) to make it more likely that a system will be adopted. At the same time, access to information legislation and human rights law may also present design opportunities \cite{Kirkham2020Using} – it is important to think more holistically about legislation, as even arcane provisions can serve as alternatives. Accordingly, there is need to have an economic model that also takes into account the wider legislative context – if a system is not able to lawfully and efficiently provide documentation, then it will probably be inadequate when used in practice.

\section{Proof of Concept Case Study – Application of our Framework}
\subsection{Scenario: People with Mobility Impairments in Europe}
In what follows, we present a carefully chosen \textbf{example }that illustrates how our system might be applied to produce a useful system for some people with a mobility impairment. 
Choosing a scenario that is likely to demonstrate a real world impact mostly relates to \textit{Issue 1} in our framework. Our scenario concerns a form of sensing that addresses surface types in the European Context. In Europe, there is a long-standing legacy of old areas, with cities including cobblestones, grass paths and legacy paving types. Compared to countries like the United States or Australia, European cities tend to have evolved organically to include a diverse variety of different path types within a small space: this is due to the fact that most cities were based on medieval structures and layouts \cite{Koch2010Die}. The advantage of European style cities is that they have relatively short path segments, so it is not as critical for street features (e.g., curb-cuts) to always be in the right place, compared to US cities that have relatively large street blocks \cite{Koch2010Die}. Moreover, the existence of relevant accessibility affordances is highly correlated with the surface type, because newer surfaces (e.g., asphalt) will normally involve the installation of accessibility features as they are replaced \cite{Hara2013Combining}. Taking a holistic view in line with \cite{Gupta2020Towards}, we focus on classifying surfaces, as classifying them into different types could substantially improve navigation for people with mobility impairments, without the difficulties of directly classifying specific street features.
\subsection{Methodology}
\subsubsection{Data Collection.} We mounted a modern smartphone via a commodity camera mount (similar to that used in \cite{Kirkham2017WheelieMap:}) on a rollator (or walker) (see \textbf{Figure 1}).  We chose a camera based approach given that inertial sensors are sensitive to individual users and have so far not shown promising performance for accessibility documentation. The smartphone was mounted at a height of 50cm at a 60° angle so that the surface could be directly observed, whilst ensuring that identifying pictures of individuals was not captured, in line with the approach of \cite{Kirkham2017WheelieMap:}. The advantage of this privacy sensitive approach is that we address an important concern in respect of \textit{Issue 5} of the framework, as a system that does not meet this requirement might not be permitted in a range of circumstances (similar to the fact that Google StreetView is effectively banned in Germany). The camera covers a horizontal swath of about 1.15 m: this exceeds (by over 25\%) the legal standard for barrier-free construction in Germany (DIN-18040) that says that the width of a path has to be 0.9 m for ease of access by wheelchair users. This field of view also minimizes the risk of capturing other objects, occlusions and other confounding information, thereby increasing the subsequent prediction quality \cite{Liu2019From}. 

The smartphone took a photograph every ~800ms, which were saved alongside their respective timestamp and GPS location. We traversed all of the available paths within a series of defined geographical regions in three cities, with a total path length of 36.22km within an area of 1.641km\textsuperscript{2}. The paths were traversed in 6 days by one person using a rollator. This approach was designed to ensure that the data collected was as naturalistic as possible while reflecting the variety of different surfaces, thus helping address an important concern associated with \textit{Issue 2}. The relatively cheap nature of the hardware (around \$100) and the fact that data collection fits with ordinary day to day activities are intended to help address the economic concerns noted as being fundamental in \textit{Issue 5}. For Bremen, the regions (see \textbf{Figure 3I}) include the city centre (areas A and B), a suburb (C and D) and a university campus area (E and F), whilst we also collected smaller areas of different cities, namely Hamburg (see \textbf{Figure 3II}) and Hannover (see \textbf{Figure 3III}) to enable further experiments.

\begin{figure}
    
    \begin{minipage}[b]{0.3\textwidth}
        \includegraphics[width=3cm]{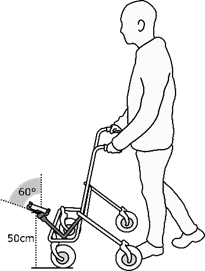}
        \caption{The data collection apparatus, including the smartphone positioning} \label{fig1}
    \end{minipage}
    \hfill
    \begin{minipage}[b]{0.7\textwidth}
        \includegraphics[height=5cm]{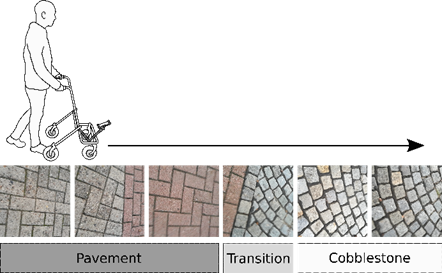}
        \caption{An illustration of how transitions were identified and labelled in the collected data} \label{fig2}
    \end{minipage}
\end{figure}

\begin{figure}
    \includegraphics[width=0.9\textwidth]{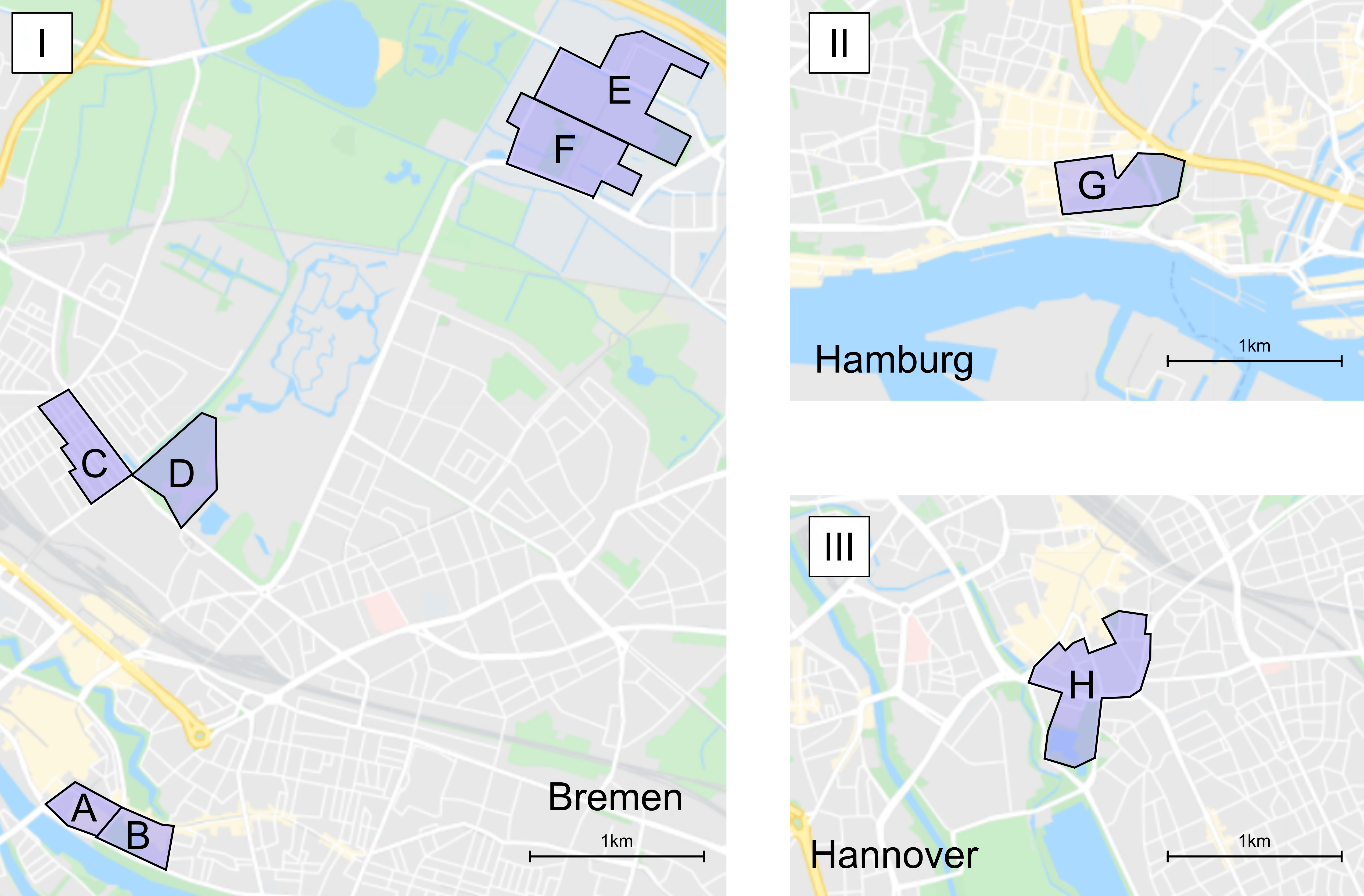}
    \caption{Map of regions within the city that were explored for our evaluation. In Figure 3(I) covering ‘Bremen’ A and B are in the city centre, C and D are in a suburb, whilst E and F are within a University campus of a major University. Figure 3(II) covers ‘Hamburg’, whilst Figure 3(III) covers ‘Hannover’.} \label{fig3}
\end{figure}

\subsubsection{Data Categorization.} A key concern was to have easily annotatable categories that relate to the real concerns of people with disabilities. To address this part of \textit{Issue 1}, the categories for the surfaces are a subset of those provided in OpenStreetMap, which is the pre-dominant open source mapping framework \cite{Haklay2008OpenStreetMap:}. The advantage of using Open-StreetMap is that these categories are widely tested and used in practice (allowing them to be easily and consistently annotated), as well as having already been con-figured to be usable by people on the ground. Taking account of the framework for considering accessibility barriers in [58], these categories were carefully curated and merged to: (i) eliminate categories that were not present in the locale (e.g., snow, ice and salt), (ii) to ensure that the boundaries between categories that are well defined (so they can be effectively annotated), merging where necessary and (iii) to ensure that only categories that are relevant to wheelchair accessibility were included. The resulting categories are listed below, with indicative examples being provided in \textbf{Figure 4}:

\paragraph{\textbf{Asphalt:}} Asphalt is an improved surface that is generally prioritised by wheelchair routing tools, due to its smooth quality (provided it is maintained) and the likelihood of other accessibility affordances being available (e.g., dropped curbs) \cite{OpenRouteService,Routino}.

\paragraph{\textbf{Cobblestones:}} Cobblestones are an old form of paving, which is inaccessible to many wheelchair users and present a severe trip hazard \cite{Kasemsuppakorn2008Data}. In practice, they are to be avoided by any routing tool, if at all possible.

\paragraph{\textbf{Ground/ Unimproved:}} This is a path that is a dirt track, perhaps including gravel. Whilst this surface type is generally traversable (albeit uncomfortably \cite{Daveler2015Participatory,Kasemsuppakorn2008Data}), there is a reduced likelihood of there being accessibility affordances, due to the unimproved nature of this path.

\paragraph{\textbf{Grass:}} This is where there is no path at all, for example where someone has been directed through a park. This is a difficult surface for most wheelchair users to traverse \cite{Daveler2015Participatory}, whilst also indicating that there no substantive path, thus meaning that accessibility affordances (for instances, markers for visually impaired people to aid navigation are less likely to be present, given the open nature of these spaces).

\paragraph{\textbf{Pavement:}} Pavement (see \textbf{Figure 2}) is an improved surface that is generally prioritised by wheelchair routing tools, due to its smooth quality (provided it is maintained) and the likelihood of other accessibility aid (e.g., dropped curbs). However, unlike asphalt, it can be an intrinsic trip hazard to wheelchair users (and ambulant people with gait or visual impairments), due to small variations in slab height \cite{Fotios2018Illuminance}.
\\

We also annotated transitions between more than one category separately (in effect making them as a sixth class), with a transition being defined as the point where more than one different surface class was simultaneously visible to the camera. This process is illustrated in \textbf{Figure 2}. More generally, the ground-truth labelling of the images was performed by the research team over a period of two working days. This could be done efficiently due to the fact that consecutive images often had the same surface (with a mean of 64.93 images being in a batch), allowing for a Panopticon \cite{Jackson2013Panopticon:} style approach to be used to speed up the annotation (but with frames, rather than videos).

\subsubsection{Testing and Training Data.} The choice of appropriate testing and training data specifically fits within\textit{ Issue 3}. We adopted two different types of leave one out approach for testing and training our approach: 

\begin{figure}
    \centering
    \includegraphics[width=0.9\textwidth]{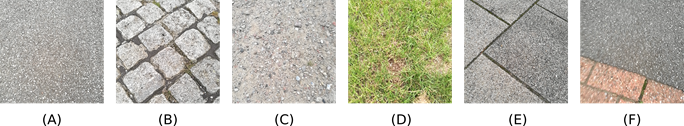}
    \caption{Exemplars of different classes, from left to right: (A) Asphalt, (B) Cobblestones, (C) Ground/Unimproved, (D) Grass, (E) Pavement and (F) Transition.}
    \label{fig4}
\end{figure}

\paragraph{\textbf{Conservative:}} This is where areas geographically close together are not trained together. In practice this means testing areas A and B together (with areas C, D, E and F being the training data) and so on for areas C and D, and E and F. This makes three areas, S1 (equal to A and B), S2 (equal to C and D) and S3 (equal to E and F). The approach here is designed to be a ‘worst-case-scenario’ and is highly likely to understate recognition performance.

\paragraph{\textbf{Leave one region out:}} In this case, we only leave one region out. For example, Region C would be trained on Regions A, B, D, E and F. This approach is designed to be in line with typical performance, if further training data were to be collected (for example in other cities).
\\

By deliberately ensuring that the same area does not appear in both the testing and training data, we avoid the error that has been made in most previous works on accessibility mapping and thus ensure our algorithm generalizes to novel examples that it was not trained on.

\subsubsection{The Machine Learning Algorithm.} Recent developments in deep learning and in particular the use of Convolutional Neural Networks has led to substantial improvements in surface texture classification \cite{Liu2019From}. Addressing\textit{ Issue 4 }requires adopting an approach which is already known to be effective for the problem at hand. We adopt one such approach, relying on the architecture of ResNet50 \cite{He2016Deep}: notably, ResNet approaches have already been used in other accessibility documentation scenarios \cite{Weld2019Deep}. This network architecture was retrained afresh on our dataset, with the only modification to this architecture being in the final layer, which was changed to have the correct number of output nodes. The images were cropped to a window of 480 by 480 pixels (the camera was positioned in portrait mode, with the bottom 160 rows being cropped). This was then interpolated to 224 by 224 pixels to be an appropriate resolution for ResNet50 implementation. We deliberately adopted a ‘motion independent approach’ to address the concern (per \textit{Issues 2 and 3}) that machine learning algorithms are not well adapted to people with disabilities (and the fact that there re-mains a lack of metrics to evaluate the extent of this problem). Specifically, to ensure that our approach was independent of the motion of the rollator, each frame was trained independently of each other: we did not use any computer vision techniques (such as optical flow) that deal with sequences of images.

\begin{table}
\caption{Summary of data collected, including class label distributions, by individual frame. Regions A to F are all in Bremen.}\label{tab1}
\centering 

\begin{tabular}{|C{0.23\textwidth}|C{0.10\textwidth}|C{0.10\textwidth}|C{0.10\textwidth}|C{0.10\textwidth}|C{0.10\textwidth}|C{0.10\textwidth}|C{0.10\textwidth}|}

\hline
\multicolumn{1}{|c|}{\rotatebox[origin=c]{90}{\textbf{Region}}} & \multicolumn{1}{|c|}{\rotatebox[origin=c]{90}{\textbf{Asphalt}}} & \multicolumn{1}{|c|}{\rotatebox[origin=c]{90}{\textbf{Cobblestone}}} & \multicolumn{1}{|c|}{\rotatebox[origin=c]{90}{\textbf{Grass}}} & \multicolumn{1}{|c|}{\rotatebox[origin=c]{90}{\textbf{ Ground/Unimproved }}} & \multicolumn{1}{|c|}{\rotatebox[origin=c]{90}{\textbf{Pavement}}} & \multicolumn{1}{|c|}{\rotatebox[origin=c]{90}{\textbf{Transition}}} & \multicolumn{1}{|c|}{\rotatebox[origin=c]{90}{\textbf{Total (in Region)}}} \\
\hline
\multicolumn{1}{|c|}{\textbf{A}} & 0 & 1656 & 0 & 0 & 930 & 632 & 3218\\
\multicolumn{1}{|c|}{\textbf{B}} & 44 & 577 & 0 & 1224 & 1696 & 423 & 3964 \\
\multicolumn{1}{|c|}{\textbf{C}} & 1017 & 47 & 0 & 0 & 3501 & 300 & 4865 \\
\multicolumn{1}{|c|}{\textbf{D}} & 78 & 132 & 662 & 4252 & 0 & 39 & 5163 \\
\multicolumn{1}{|c|}{\textbf{E}} & 1500 & 476 & 571 & 288 & 1940 & 161 & 4936 \\
\multicolumn{1}{|c|}{\textbf{F}} & 1249 & 785 & 807 & 730 & 2677 & 192 & 6440 \\
\multicolumn{1}{|c|}{\textbf{G (Hamburg)}} & 619 & 563 & 381 & 572 & 3034 & 227 & 5396 \\
\multicolumn{1}{|c|}{\textbf{H (Hannover)}} & 1136 & 1090 & 333 & 957 & 3612 & 211 & 7339 \\
\multicolumn{1}{|c|}{\textbf{Total}} & 5643 & 5326 & 2754 & 8023 & 17390 & 2185 & 41321 \\ 
\hline
\end{tabular}
\end{table}

\subsection{Results}
Based on the real-world application of our framework we got results referring to the data collection and categorization but also for the performance of the trained system itself. The latter can be considered in relation to leave one out frames and conservative frames whereby both are investigated framewise and “streetwise”. The first three results subsections focus on Bremen, whilst our final exercise compares the three cities.

\subsubsection{Data Collection and Distribution.} The eight regions A-H were all annotated, producing a total of 41321 individual frames. As can be seen in \textbf{Table 1}, there is skew in the class distribution, with certain types of surface being more common than others. The most common surface type was the pavement, followed by unimproved surfaces. Two of our analyses focus on within city performance and thus just 'Bremen', whilst we look at cross-city performance in Section \emph{Cross-City Performance}.
\begin{figure}
    \includegraphics[width=0.5\textwidth]{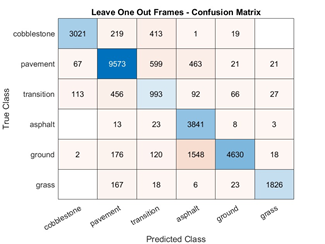}
    \includegraphics[width=0.5\textwidth]{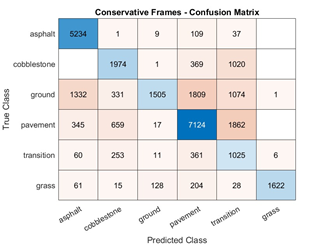}
    \caption{. Confusion Matrices for both ‘Conservative’ and ‘Leave one region out’ on a \emph{Frame} basis.} \label{fig5}
\end{figure}

\begin{figure}
    \includegraphics[width=0.5\textwidth]{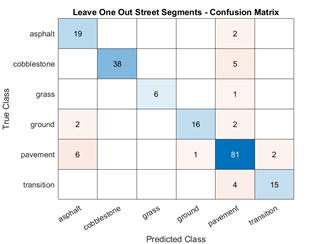}
    \includegraphics[width=0.5\textwidth]{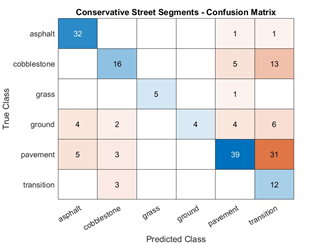}
    \caption{Confusion Matrices for both ‘Conservative’ and ‘Leave one region out’ on a \emph{Street} basis.} \label{fig6}
\end{figure}

\subsubsection{Framewise performance.} We report the raw performance per frame based on both approaches, conservative and leave one region out based on 'Bremen'. The conservative performance was a mean F1-score \cite{Hammerla2013On} of 0.621, whilst the leave-one-region-out mean F1-score was 0.802. The more detailed performance can be seen in the con-fusion matrices, \textbf{Figure 5}, which shows that most relevant categories (asides transitions) were easily distinguishable, but with reduced performance in respect of certain categories, for example ground was regularly confused with asphalt and in the more conservative approach, with other categories.

\subsubsection{Aggregated ‘Streetwise’ Performance.} In practice, the concern is to identify whether streets contain problematic elements, rather than identifying each frame. Following \cite{Ward2011Performance}, we therefore report results on a streetwise level using an aggregation process, addressing an important concern under \textit{Issue 2 }which hitherto has not been addressed in the context of accessible mapping. The conservative performance was a mean F1-score of 0.593, whilst the leave one region out mean F1-score was 0.871. The more detailed performance can be seen in the confusion matrices, \textbf{Figure 6}, which shows that most relevant categories (asides transitions) were easily distinguishable in the leave one region out case. In respect of the more likely leave one out approach, the effect is to substantially increase the performance (there was a slight decrease in the other case, but this could be statistical noise), whilst reporting the results in a manner that more directly relates to real world performance.

\begin{figure}
    \includegraphics[trim=4cm 0cm 4cm 0cm,clip,width=1\textwidth]{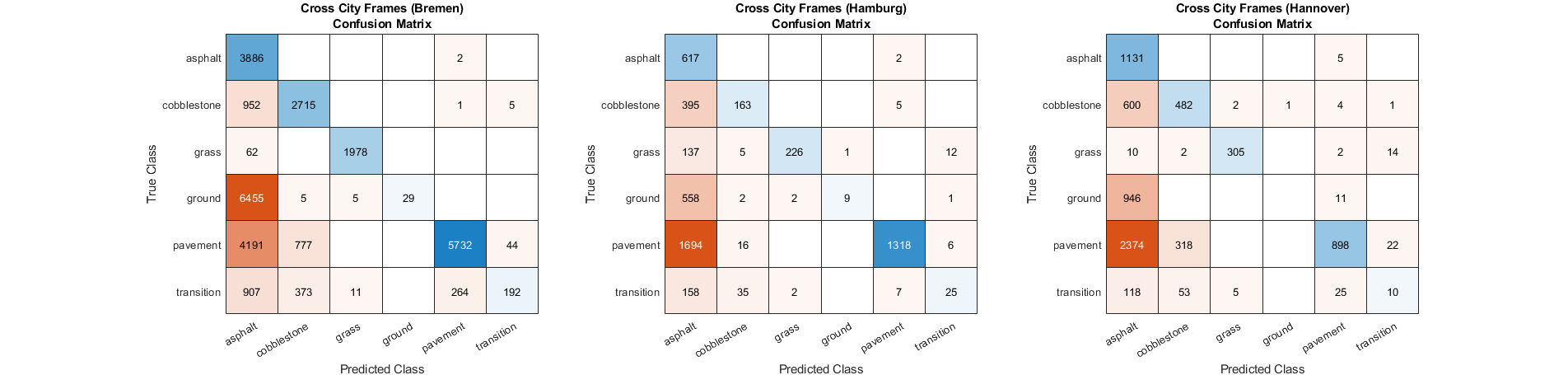}
    \includegraphics[trim=4cm 0cm 4cm 0cm,clip,width=1\textwidth]{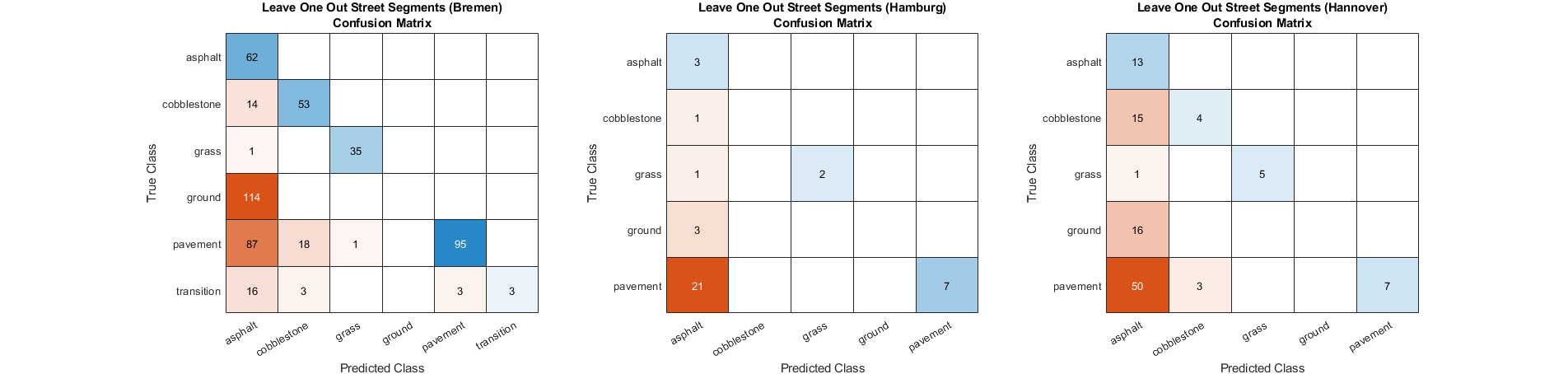}
    \caption{Confusion Matrices for both ‘Conservative’ and ‘Leave one region out’ on a \emph{Cross-City} basis.} \label{fig7}
\end{figure}

\subsubsection{Cross-City Performance.} We present the performance of our approach where a system is only trained on data from other cities (so for example we would train on ‘Hamburg’ and ‘Hannover’ and test on ‘Bremen’). This performance is poor, making it clear that some training data within a city is necessary. In respect of frame-wise performance, we obtain a mean F1-score of 0.495 for 'Bremen', 0.378 for 'Hamburg' and 0.378 for Hannover (Hannover was slightly higher, but rounding to three decimal places, we get the same value as for Hamburg). For streetwise performance (following the same process in \emph{Aggregated ‘Streetwise’ Performance}), we obtain a mean F1-score of 0.487 for 'Bremen', 0.278 for 'Hamburg' and 0.333 for Hannover. In \textbf{Figure 7}, we provide the confusion matrices illustrating more detailed performance.

\section{Discussion}
\subsection{Real-World Surface Recognition Performance}
Our framework was helpful in designing a practical system that obtained a high real-world performance. It is apparent from the results above that is possible to reliably recognize different surface types using our approach. The more conservative approach performed worse, as one would expect, partially because of the reduced volume of training data: however, even in this relatively extreme scenario, the performance would still improve upon existing tracking systems (noting that none of these is any better than pedestrian routing tools \cite{Tannert2019Analyzing}) and is substantially better than choosing a given route by chance. 

The less conservative and more realistic (taking into account the considerations in \cite{Hammerla2015Let's}) approach of leave one region out yielded stronger performance and if this were to be real world performance, would lead to appreciably (rather than technically) better approaches for routing people with disabilities. Moreover, this approach is promising for other goals in addition to making more effective routing tools: for example, this approach would be sufficiently accurate to estimate the accessibility of a given city, or the distance travelled by someone on a rough surface. 

In a real-world context, the decision to be taken really is a two-class problem – whether or not a path segment is accessible. As such, the general f-ratio across our six classes actually considerably understates performance in a navigation scenario. A more realistic approach would ignore the transitions (which could normally be identified using GPS and existing maps) and split the remaining classes. For example, doing this with asphalt and pavement being considered to be suitable surfaces and the rest be deemed unsuitable would lead to an f-ratio of 0.952 per street segment, if applied to the leave one out scenario. This would be adequate to produce an accuracy on a route of over 82\%, if each route was assumed to be 4 path segments long and the accessibility of each segment independent from one another. However, this assumes that training data is collected from the same city in a sufficient volume to be adequate: training on one city and using it for another was shown to be infeasible (see Section Cross-City Performance) above, thus emphasizing the need for data to be quickly obtainable.  

\subsection{The Importance of Frameworks: Accessibility mapping as a strategy problem}
Our approach evidences the advantages of a camera mounted to a mobility aid (and combined with GPS) as being a model for accessibility barrier detection and mapping. By carefully considering the framework questions we set out in designing a sensing system in this space, we were able to adapt existing machine learning approaches to create a practical and realistic approach for surface class detection. From a practical point of view, our approach is fully automatic, can be done by attaching a simple device to any mobility aid (in part due to the motion independent design) and training data can be annotated at speed relative to motion-based systems. Indeed, training such a system is very economic, given the data can be collected and annotated in a matter of a few days, meaning it would be feasible to perform this exercise for each individual city (thus greatly reducing the practical impact of the relatively low 'cross-city' performance, i.e. where a system trained on one city is used in another). We also fully address important legal issues, especially privacy, which can be a practical challenge (e.g., in Germany, there remain heavy restrictions that would otherwise apply to street photography per the “Bundesverband Informationswirtschaft, Telekommunikation und neue Medien” \cite{Machine}).  At the same time, this work can be directly used in a specific setting: for instance, the documentation generated by the surface detection described in our case study might in turn be used in a navigation tool to reduce risk of falls in elderly people by reducing the probability they will need to follow a route containing an unimproved surface. 

In effect, this work treats accessibility mapping as primarily a strategy problem, rather than a technical machine learning problem. It reflects the fact that there is a ‘bigger’ and more complex picture that is perhaps unique to accessibility sensing. In articulating these concerns and providing an exemplar of how to address them, we offer a clear road forwards in offering an automated approach, which we hope help resolve the wider challenge of inaccessibility in the built environment. Our approach of developing a set of practical principles that could be adopted more widely (rather than just focusing on one system) offers a means for addressing wider concerns about AI fairness with respect to people with disabilities, with the principles we have identified in respect of accessible navigation serving as a potential starting point for frameworks that apply to other settings.

\subsection{Real world performance}
Our work is the only one so far to adopt a real-world approach towards assessment of performance, including in respect of appropriate metrics in line with \cite{Ward2011Performance}. This is an important consideration that should be taken up in this field: presently it is not possible to understand the performance of existing systems and thus to understand progress. The implications of this for our work is that we can reliably expect our results will generalize and thus have clearly illustrated the benefits of our camera centric approach for accessibility mapping. We expect (and sincerely hope) that a renewed focus on performance metrics that more directly relate to the problem of accessible navigation will be something given an increased emphasis in this sphere. Going forwards, this would allow us to directly determine whether our work is more effective than others, as presently it is very difficult to compare real-world navigation performance as existing papers do not report ‘street segment’ performance. 

\subsection{Limitations}
This work has clearly shown the advantage of a given strategy in documenting accessibility barriers in the built environment, yet we still have some important limitations. First, whilst we identify appropriate surface types, this does not ad-dress the need to identify broken surfaces (e.g., cracks in the pavement) and other trip hazards. Second, the dataset only contains pictures from a six-day period without influence of rain or seasons like snow. The third concern is motion-blur, which occurs where there are quick movements to pull the apparatus up a curb: addressing this may increase performance (e.g., using optical flow to give less weight to frames that include this). Fourth, future work may have a larger dataset available, potentially increasing performance further, likewise with any approach that combines multiple sources of information, as opposed purely our on-ground assessment process. Finally, we note that our framework itself should evolve, as we develop our knowledge of fairer machine learning approaches for people with disabilities, and better accessibility documentation systems.

\section{Conclusion and Future Work}
We have presented a new approach towards documenting accessibility barriers in the built environment and measuring the performance of a recognition system in this space, focusing on the practical problems faced by people with disabilities. This approach is pragmatic, reflecting that the problem of concern is an intersection of HCI concerns and applied machine learning. To explore this approach, we presented a case study, which implemented an instance of our framework in a fully automated pipeline that operates effectively in the navigation context. Future work could explore more instances of this approach, thereby helping to concrete this strategy for documenting barriers going forwards, as well as translating it to a real-world system.

%
%
%
\bibliographystyle{splncs04}
\bibliography{references}

\end{document}